# Control of domain wall motion at vertically etched nanotrench in ferromagnetic nanowires

Kulothungasagaran Narayanapillai and Hyunsoo Yang[a)]

*Department of Electrical and Computer Engineering, National University of Singapore, 117576, Singapore*

We study field-induced domain wall motion in permalloy nanowires with vertically etched nanotrench pinning site. Micromagnetic simulations and electrical measurements are employed to characterize the pinning potential at the nanotrench. It is found that the potential profile for a transverse wall significantly differs from that of a vortex wall, and there is a correlation between the pinning strength and the potential profile. Reliable domain wall pinning and depinning is experimentally observed from a nanotrench in permalloy nanowires. This demonstrates the suitability of the proposed nanotrench pinning sites for domain wall device applications.

[a)] e-mail address: eleyang@nus.edu.sg

Domain wall (DW) based devices have been proposed as a promising solution for future high density storage and logic devices[1-6]. Implementing these devices requires precise and reliable control of DWs which can be achieved through pinning centers. Automotion happens in ideal nanowires[7], however, in reality, lithographic imperfections in nanowires results in natural pinning sites which are inherently hard to control. Controllable pinning sites can be introduced by engineering artificially structured variations into the geometry along the nanowires, which are known as "notches". Complex DWs are formed at these notches and its depinning behavior is very sensitive to the initial state of the DW, its structure, and chirality as well as the excitation method[7-10]. Even the stochastic behavior of the pinning and depinning process can be controlled by varying the notch dimensions[11]. Static and dynamic pinning strengths for a DW at the pinning sites have also shown deviations[12]. All these results show that the control of DWs depends predominantly on the notch profile.

A few mechanisms are typically used to pin and control a DW in ferromagnetic nanowires of in-plane anisotropy systems. Ion irradiation is an approach in which a portion of the nanowire is implanted with ions to soften the magnetic properties, thereby creating a pinning site[13,14]. The common approach is to introduce lateral constrictions along the nanowire, as a result, giving rise to a notch[15] or alternatively a lateral protrusion known as anti-notch[12]. However, in nanowires, it is challenging to precisely control the lateral dimensions at the nano-scale due to the limit of modern lithography techniques.

In this letter, we demonstrate an alternative approach to control the notch profile vertically by removing a rectangular shaped portion of the magnetic material along the nanowire (hereafter referred to as nanotrench). It enables us to utilize the advantage of controlling the vertical etching depth accurately down to a few monolayers with ion milling. We study the field-induced pinning and depinning of DWs, from the nano-trench pinning site, in the permalloy nanowires. Micromagnetic simulations and electrical measurements are employed to characterize the potential strength of these pinning sites. The pinning strength linearly increases with the depth of the nanotrench for transverse and vortex DWs. Above a certain length of the nanotrench, the depinning strength begins to saturate. The stochastic nature of DW generation and depinning is also presented in a nanotrench.

Micromagnetic simulations of the depinning studies are performed using the object oriented micromagnetic framework (OOMMF). Two different dimensions of nanowires are utilized for studying transverse (a width of 100 nm and thickness of 10 nm) and vortex (a width of 200 nm and thickness of 40 nm) DWs. Cell dimensions of $4\times4\times2$ $nm^3$ and $4\times4\times5$ $nm^3$ have been used, for the transverse and vortex case, respectively. A saturation magnetization of $M_S=8.6\times10^5$ A/m, exchange constant of $A=13\times10^{-12}$ J/m, and an anisotropy constant of $K = 0$ are assumed. The simulations were performed at the quasi-static regime and the Gilbert damping parameter ($\alpha$) is set to 0.5 to improve the speed of the simulations.

A nanotrench is placed at the center along the nanowire as shown in Fig. 1(a). The DWs are initially located at the right edge of the nanotrench and then released to relax for several nanoseconds to form an energetically favorable and stable structure in each simulation. Examples of a similarly initialized transverse and a vortex DW at the nanotrench are shown in Fig. 1(b) and (c), respectively. In both cases, the length of the nanotrench (LN) is 240 nm, while the depth (DN) is 6 and 20 nm, respectively, for transverse and vortex cases. The position of the nanotrench is highlighted using a dark shade. During the relaxation process, the vortex DW moves outwards of the nanotrench and is stabilized, while the transverse DW moves towards the center of the nanotrench.

The effect of varying the nanotrench dimensions on the pinning and depinning of both types of DWs have also been studied. The length (LN) and the depth (DN) of the nanotrench are gradually varied. For the case of a transverse wall, when the depth is increased from 2 to 6 nm, the depinning field strength increases almost linearly as shown in Fig. 2(a). This trend remains the same for all different lengths investigated. However, when the length is increased from 20 to 240 nm, it is observed that the pinning field increases linearly and, subsequently, saturates as shown in Fig. 2(b). For the case of 6 nm deep nanotrench, the depinning field saturates at a length of 100 nm. A similar trend is also seen for the vortex DWs which have been plotted in Fig. 2(c) and (d). The depinning field increases with the depth and length of the nanotrench. However, it is evident that irrespective of the depth of the nanotrench, the depinning strength profile saturates around a length of ~100 nm for both cases of transverse and vortex DWs. This saturation behavior can be understood by the energy landscape of the pinning sites as we discuss below.

In order to provide a more quantitative understanding of the depinning behavior, the potential landscape for the DW states are calculated by micromagnetic simulations. The geometrical variations along the nanowire generate an energy landscape that a DW experiences while traversing through the wire. The change in the potential profile reflects the interaction between the spin structure of DW and the pinning site. In order to understand the energy landscape of the nanotrench profile, energy terms like demagnetization and exchange energy are taken into consideration. A DW is initially placed on the left side, at a distance of 1.5 μm away from the center of the nanotrench. A constant magnetic field is applied along the $+x$-direction to drive the DW towards the right end of the nanowire, thereby, passing through the nanotrench at the center of the nanowire. The absence of the anisotropy energy term makes the energy equation converge to $E_{Tot}$=$E_{Dem}$+$E_{Ex}$, where $E_{Dem}$ is the demagnetization energy and $E_{Ex}$ is the exchange energy.

The contributions of each energy term in the system for a transverse DW (LN=240 nm; DN=6 nm) is plotted in Fig. 3(a) with respect to the DW position. The energy is normalized with respect to the total energy ($E_{Tot}$). The center of the nanotrench is set to zero on the *x*-axis. The total energy of the system is locally reduced forming a potential well around the center of the nanotrench in case of the transverse DW as shown in Fig. 3(a). The demagnetization energy contributes 90%, while the exchange energy is ~10% of the total energy contribution outside the potential well. The exchange energy increases while entering into the nanotrench area, as a result, providing resistance to the moving DW, indicated by a small peak in $E_{EX}$ at -0.19 μm in Fig. 3(a). The energy landscape of the vortex wall (LN=240 nm; DN=15 nm) are shown in Fig. 3(b). It should be noted that the interactions generated at the nanotrench edge by a larger vortex DW structure will make the energy landscape different from the transverse DW case. The transverse DW pins at the center of the pinning site, whereas the vortex DW is repelled away from the center of the nanotrench, but pins at either edges of the nanotrench due to a dual-dip energy profile. This phenomenon is also observed in the conventional constriction type notches, where the vortex DW has to realign its spin structure at the expense of increasing energy terms while passing through the notch[9,16]. However, the contributions from the demagnetization energy and exchange energy remain around 90% and 10% outside the potential well, respectively, which is similar to the transverse DW case. Similar energy contributions have been reported for pinning sites defined by ion implantation[13].

The total energy is plotted in Fig. 3(c) and (d) for both transverse (DN=6 nm) and vortex (DN=15 nm) walls for various lengths (LN=40 to 240 nm) of nanotrenches in order to achieve a better understanding of the energy landscape at the pinning site. It can be inferred from both cases that the shape of the energy profile is almost conserved. Furthermore, the depth of the dip in the energy profile increases with increasing the length of the nanotrench. The curvature as well as the depth of the energy landscape determines the pinning strength. The insets of Fig. 3(c) and (d) show the drop in the energy profile, $\Delta E_{\mathrm{Tot}}$ with respect to the energy at –0.5 μm which follows the same trend as that of the depinning strength discussed in Fig. 2(b) and 2(d) for both types of the DWs. For the vortex DW case, the drop in energy on both sides of the dip (labeled left and right) shows a very similar behavior as shown in the inset of Fig. 3(d).

The proposed pinning sites are experimentally verified in permalloy nanowires. Thin films with the stack structure of substrate/Ta (3 nm)/$Ni_{81}Fe_{19}$ (30 nm)/Ta (3 nm)/Ru (2 nm) are deposited in a dc-magnetron sputter tool at a base pressure of $1 \times 10^{-9}$ Torr. Sub-micrometer wires are patterned by electron beam lithography (EBL) followed by Ar ion milling. The measurement contact pads are defined by EBL, and followed by deposition and a lift-off process. The top 2 nm of the nanowire was partially etched before depositing the contact pads to provide ohmic contacts between the nanowire and the contact pads. Finally, the nanotrench is defined using an etch mask designed by EBL followed by Ar ion milling to remove a portion of the nanowire in order to form the required vertical nanotrench. Figure 4(a) shows a scanning electron micrograph with the measurement schematics along with the nanotrench highlighted in red color within the nanowire. The width of the nanowire is 650 nm and the length is 12 μm.

Anisotropic magnetoresistance (AMR) is a suitable choice for DW detection. It reduces the resistance of the nanowires due to the presence of a DW. The following sequence is employed for DW generation and detection. First, a saturation magnetic field, $H_{\mathrm{SAT}}$ = 1 kOe, is applied in the +$x$-direction and reduced to zero. Then the saturation resistance, $R_{\mathrm{SAT}}$ is measured with a dc current of 30 μA applied across $A_1B_1$ contacts, which is ~145.80 Ω. Secondly, a short pulse is applied across $A_1A_2$ contacts to generate a DW by utilizing the Oersted field generation method[17] and simultaneously a constant assist field of 30 Oe, $H_{\mathrm{ASSIST}}$ is applied in the $-x$-direction to push the DW to the nanotrench. Then, the resistance ($R_{\mathrm{I}}$) across $A_1B_1$ contacts is measured again at zero fields. The difference between two resistance values ($= R_{\mathrm{SAT}} - R_{\mathrm{I}}$) is associated with the DW resistance ($R_{\mathrm{DW}}$). Subsequently, 1 kOe is applied along the $-x$-direction to remove any effects from remanence. The above process is repeated to gain a statistical distribution.

The histogram of the DW generation process is shown in Fig. 4(b). The three different DW resistances ($R_{\mathrm{DW}}$ = -0.16, -0.13, and -0.10 Ω) can be explained by the existence of transverse and vortex DWs with different chirality at the nanotrench. The high occurrence of two types of DWs around -0.16 and -0.13 Ω could be due to the anticlockwise and clockwise vortex DWs, while the relatively small occurrence around -0.10 Ω can be attributed to the transverse DW configurations[18,19]. The depinning strength of the pinned DWs at the nanotrench is also investigated. After a DW is pinned at the nanotrench as discussed earlier, the magnetic field increases in steps of 2 Oe in the $-x$-direction and a representative depinning profile is shown in Fig. 4(c). As shown in Fig. 4(c), when the DW is removed from the nanotrench and moves out of the $A_1B_1$ portion, the resistance reaches the $R_{\mathrm{SAT}}$ value (145.80 Ω). The depinning strength depends on the DW type and chirality. From the histogram plot shown in Fig. 4(d), we can see that the DW depinning is distributed. This could be understood by the presence of different DW types generated during the DW generation process and stochastic behavior of the depinning process[20,21].

In summary, we have demonstrated DW wall pinning and depinning in the proposed vertical nanotrench site. The micromagnetic simulations show that the depinning strength can be effectively controlled by the proper selection of nanotrench dimensions. Different shapes of the potential profile are observed for transverse and vortex type DWs. In permalloy nanowires with nanotrench pinning sites both types of DWs have been experimentally shown to exist. Reliable pinning and depinning behaviors from a vertical nanotrench are observed. Compared to the lateral constrictions, our proposed method has a higher precision in defining the dimensions of the pinning sites in the sub-nanoscale.

References

[1] M. Hayashi, L. Thomas, R. Moriya, C. Rettner, and S. S. P. Parkin, Science **320**, 209 (2008).

[2] D. A. Allwood, G. Xiong, C. C. Faulkner, D. Atkinson, D. Petit, and R. P. Cowburn, Science **309**, 1688 (2005).

[3] S. S. P. Parkin, M. Hayashi, and L. Thomas, Science **320**, 190 (2008).

[4] A. Yamaguchi, T. Ono, S. Nasu, K. Miyake, K. Mibu, and T. Shinjo, Phys. Rev. Lett. **92**, 077205 (2004).

[5] D. A. Allwood, G. Xiong, and R. P. Cowburn, Appl. Phys. Lett. **85**, 2848 (2004).

[6] T. Ono, H. Miyajima, K. Shigeto, K. Mibu, N. Hosoito, and T. Shinjo, Science **284**, 468 (1999).

[7] M. Jamali, K. J. Lee, and H. Yang, New J. Phys. **14**, 033010 (2012).

[8] M. Hayashi, L. Thomas, C. Rettner, R. Moriya, X. Jiang, and S. S. P. Parkin, Phys. Rev. Lett. **97**, 207205 (2006).

[9] L. K. Bogart, D. Atkinson, K. O'Shea, D. McGrouther, and S. McVitie, Phys. Rev. B. **79**, 054414 (2009).

[10] M. Jamali, H. Yang, and K. J. Lee, Appl. Phys. Lett. **96**, 242501 (2010).

[11] M. Y. Im, L. Bocklage, G. Meier, and P. Fischer, J. Phys.: Condens. Matter. **24**, 024203 (2012).

[12] A. Kunz and J. D. Priem, IEEE. T. Magn. **46**, 1559 (2010).

[13] A. Vogel, S. Wintz, T. Gerhardt, L. Bocklage, T. Strache, M. Y. Im, P. Fischer, J. Fassbender, J. McCord, and G. Meier, Appl. Phys. Lett. **98**, 202501 (2011).

[14] M. A. Basith, S. McVitie, D. McGrouther, and J. N. Chapman, Appl. Phys. Lett. **100**, 232402 (2012).

[15] M. Klaui, C. A. F. Vaz, J. Rothman, J. A. C. Bland, W. Wernsdorfer, G. Faini, and E. Cambril, Phys. Rev. Lett. **90**, 097202 (2003).

[16] M. Klaui, J. Phys.: Condens. Matter. **20**, 313001 (2008).

[17] M. Hayashi, L. Thomas, C. Rettner, R. Moriya, and S. S. P. Parkin, Nat. Phys. **3**, 21 (2007).

[18] F. U. Stein, L. Bocklage, T. Matsuyama, and G. Meier, Appl. Phys. Lett. **100**, 192403 (2012).

[19] M. Munoz and J. L. Prieto, Nat. Commun. **2**, 562 (2011).

[20] G. Meier, M. Bolte, R. Eiselt, B. Kruger, D. H. Kim, and P. Fischer, Phys. Rev. Lett. **98**, 187202 (2007).

[21] X. Jiang, L. Thomas, R. Moriya, M. Hayashi, B. Bergman, C. Rettner, and S. S. P. Parkin, Nat. Commun. **1**, 25 (2010).

Figure captions

Figure 1. (a) Schematics of a nanowire with a nanotrench. Simulated a transverse (b) and vortex (c) DW at the nanotrench.

Figure 2. Dependence of depinning field with the depth (a) and the length (b) of nanotrench for a transverse wall. (c) and (d) show the dependence of depinning field for vortex walls.

Figure 3. Energy profiles with respect to DW position for a transverse (a) and a vortex (b) wall. (c) and (d) show the total energy for transverse and vortex walls for various lengths of nanotrenches. The drop in the energy profile ($\Delta E_{\mathrm{Tot}}$) is plotted in the insets for respective DW types.

Figure 4. (a) Scanning electron micrograph image of the device with the measurement schematics. (b) The histogram plot of the generated DW resistance (c) A typical depinning profile of a DW from a pinning site. (d) The histogram plot of depinning fields.

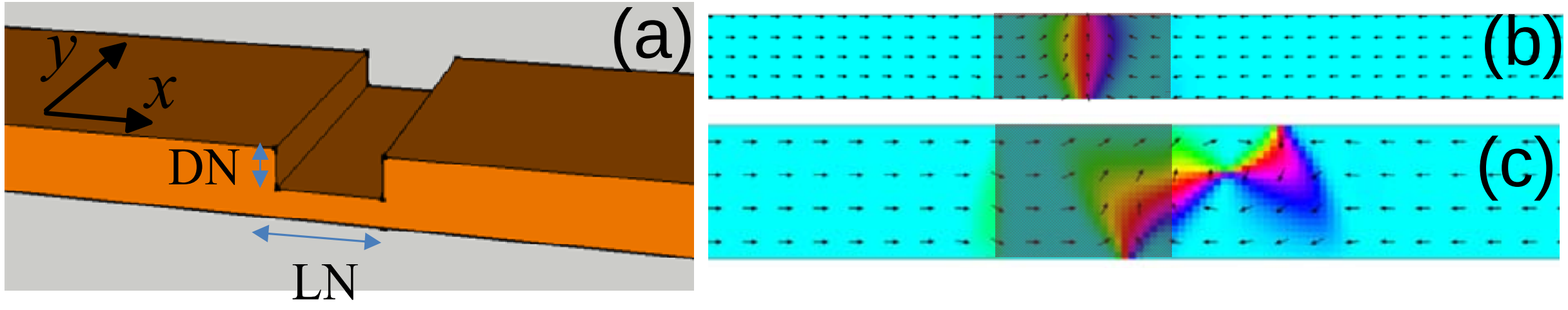

Figure 1

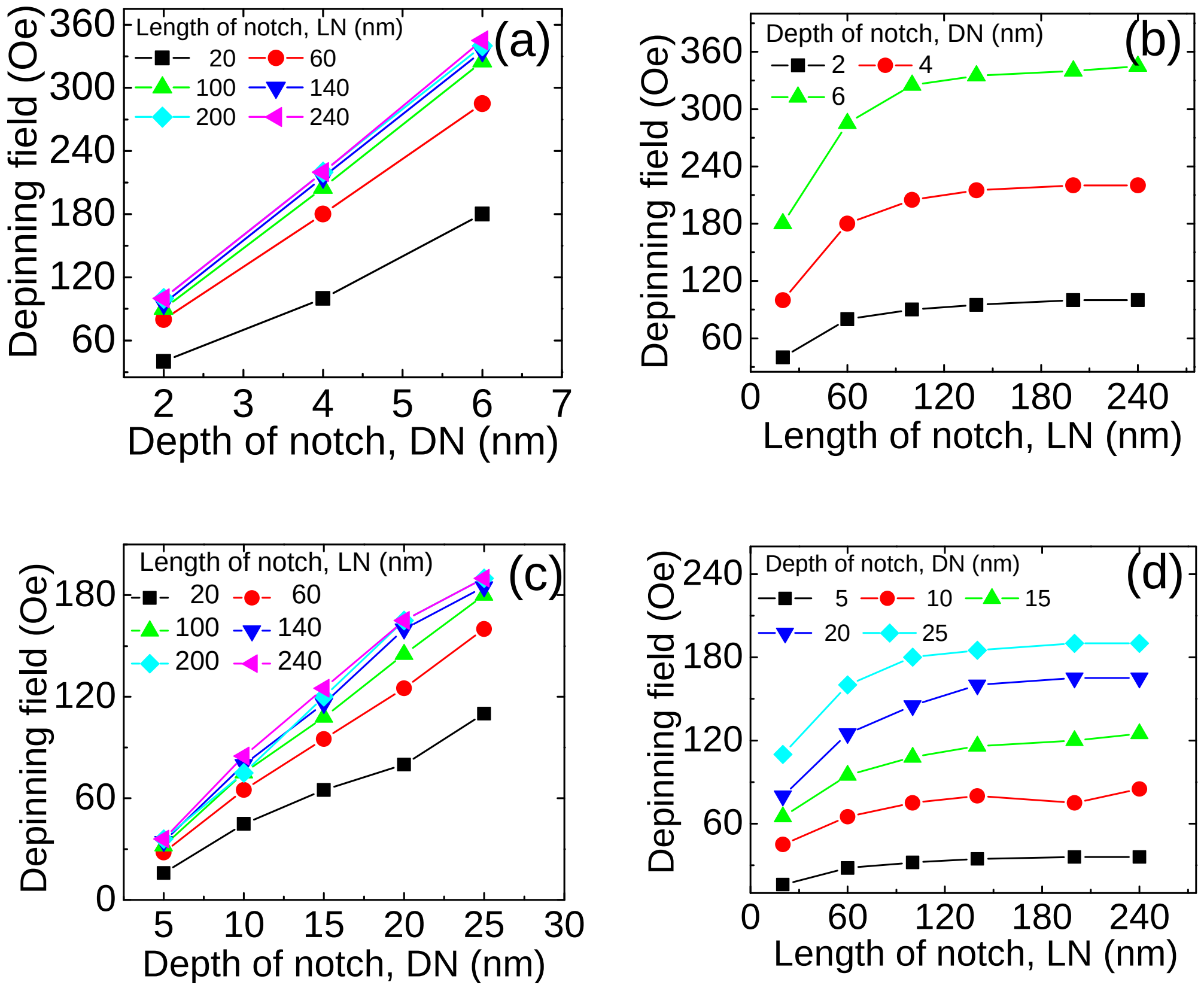

Figure 2

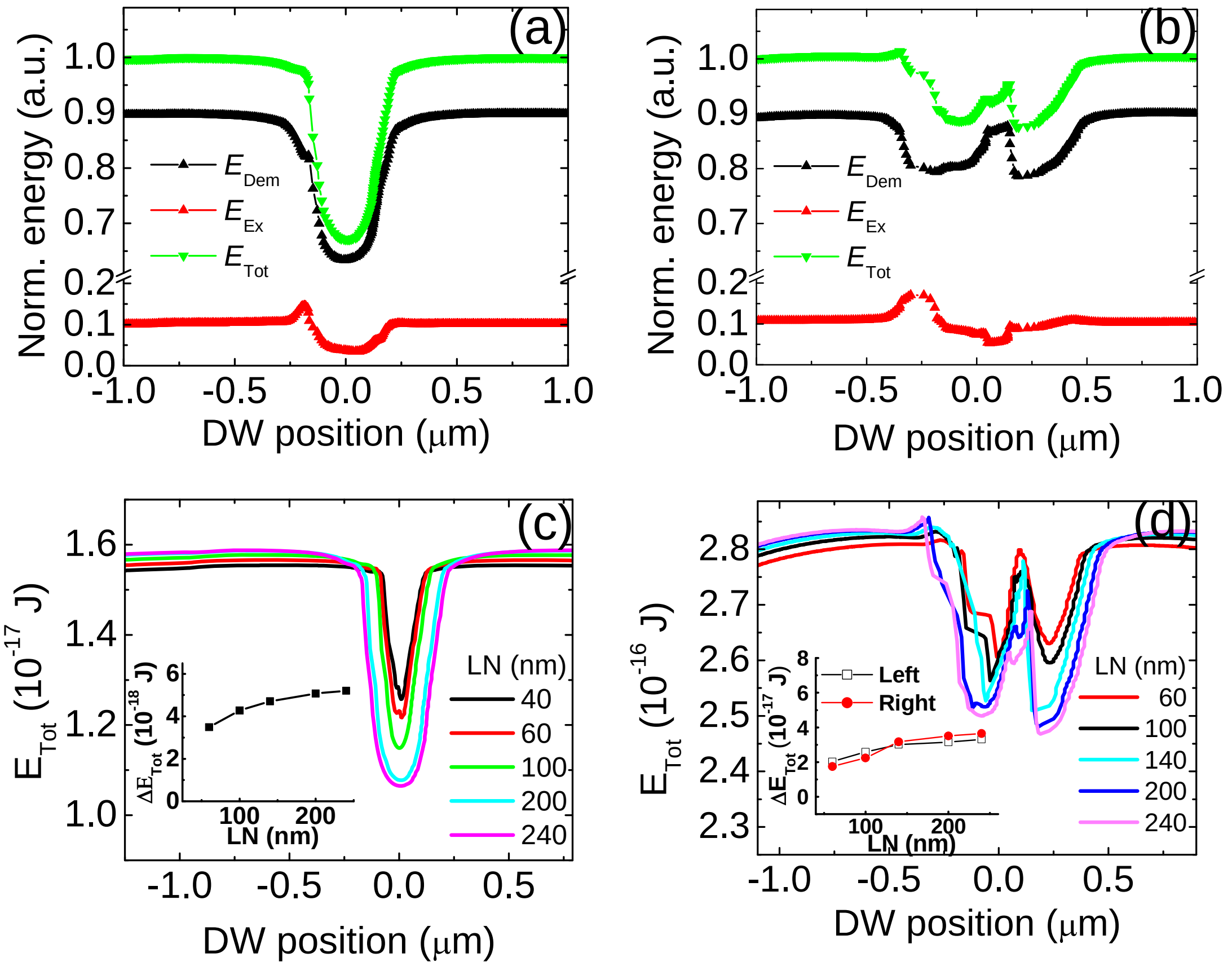

Figure 3

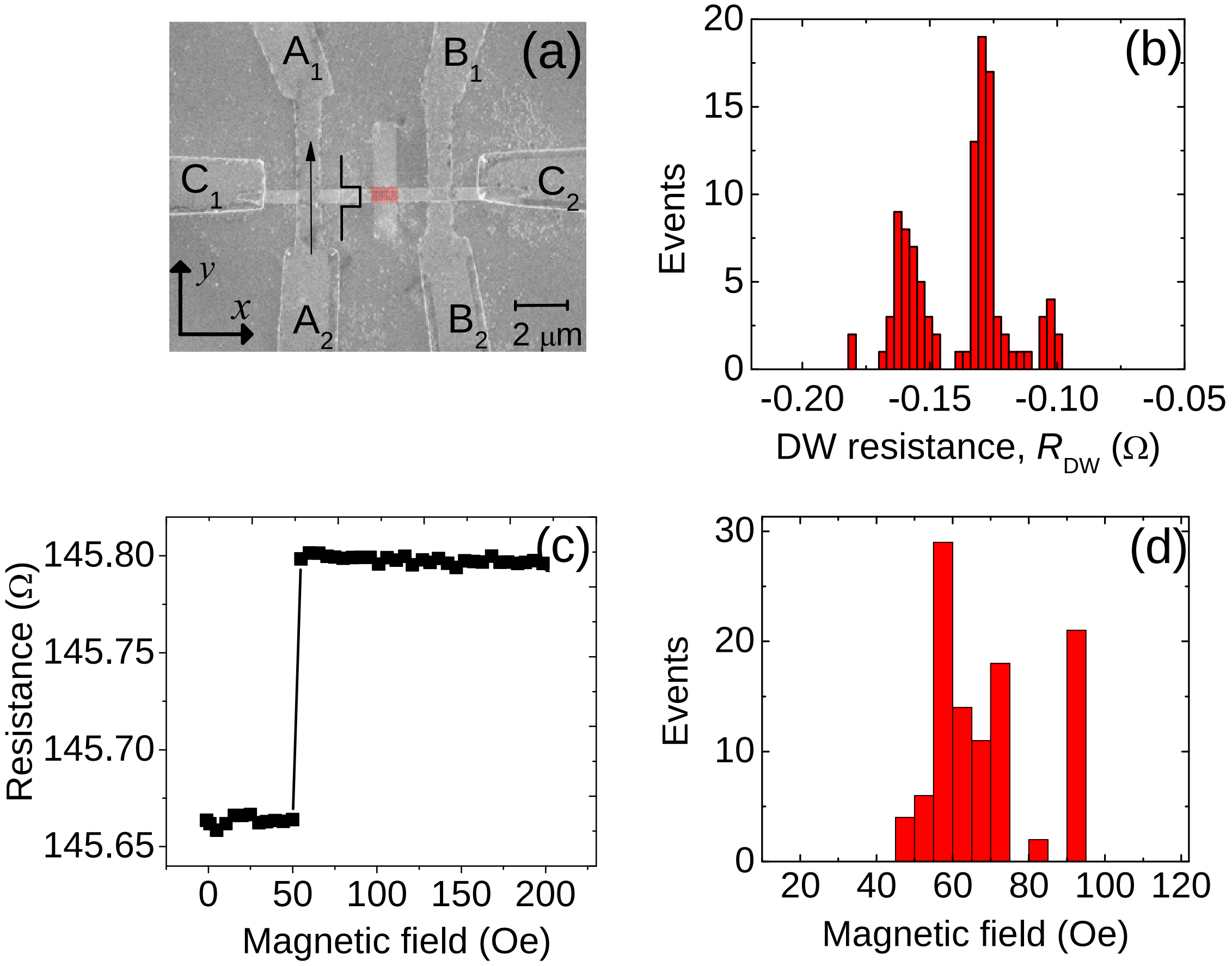

Figure 4